\documentclass[preprint,showpacs,preprintnumbers,amsmath,amssymb,nofootinbib]{revtex4-2}

\usepackage{graphicx}
\usepackage{dcolumn}
\usepackage{bm}
\usepackage{color}
\usepackage{rotating}
\usepackage{natbib}
\usepackage{ulem}

\setcitestyle{super}

\begin{document}
\newcommand\blfootnote[1]{%
	\begingroup
	\renewcommand\thefootnote{}\footnote{#1}%
	\addtocounter{footnote}{-1}%
	\endgroup}

 \newcommand{\FigCap}[1]{\textbf{#1}}	

\title{Evidence for an odd-parity nematic phase above the charge density wave transition in kagome metal CsV$_3$Sb$_5$}

\author{T.\,Asaba$^{1,\dagger}$}
\author{A.\,Onishi$^{2,\dagger}$}
\author{Y.\,Kageyama$^{2,\dagger}$}
\author{T.\,Kiyosue$^1$}
\author{K.\,Ohtsuka$^1$}
\author{S.\,Suetsugu$^1$}
\author{Y.\,Kohsaka$^1$}
\author{T.\,Gaggl$^1$}
\author{Y.\,Kasahara$^1$} 
\author{H.\,Murayama$^{1,3}$}
\author{K.\,Hashimoto$^2$} 
\author{R.\,Tazai$^{4,5}$}
\author{H.\,Kontani$^4$}
\author{B.\,R.\,Ortiz$^6$}
\author{S.\,D.\,Wilson$^6$}
\author{Q.\,Li$^7$}
\author{H.-H.\,Wen$^7$}
\author{T.\,Shibauchi$^{2,*}$}
\author{Y.\,Matsuda$^{1,*}$}

\affiliation{$^1$Department of Physics, Kyoto University, Kyoto 606-8502 Japan}
\affiliation{$^2$Department of Advanced Materials Science, University of Tokyo, Chiba 277-8561, Japan} 
\affiliation{$^3$RIKEN Center for Emergent Matter Science, Wako, Saitama 351-0198, Japan}
\affiliation{$^4$Department of Physics, Nagoya University, Furo-cho, Nagoya 464-8602, Japan}
\affiliation{$^5$Yukawa Institute for Theoretical Physics, Kyoto University, Kyoto 606-8502, Japan}
\affiliation{$^6$Materials Department, University of California Santa Barbara, Santa Barbara, California 93106, USA}
\affiliation{$^7$National Laboratory of Solid State Microstructures and Department of Physics, Nanjing University, Nanjing 210093, China}

\blfootnote{$^{\dagger}$ denotes equal contribution.}
\blfootnote{$^*$ denotes corresponding authors.}

\maketitle

{\bf The quest for fascinating quantum states arising from the interplay between correlation, frustration, and topology is at the forefront of condensed-matter physics. Recently discovered nonmagnetic kagome metals $\bm{A}$V$\bm{_3}$Sb$\bm{_5}$ ($\bm{A=}$ K, Cs, Rb) with charge density wave (CDW) and superconducting instabilities may host such exotic states. Here we report that an odd electronic nematic state emerges above the CDW transition temperature ($\bm{T_{\rm CDW}=94}$\,K) in CsV$\bm{_3}$Sb$\bm{_5}$. High-resolution torque measurements reveal a distinct twofold in-plane magnetic anisotropy that breaks the crystal rotational symmetry below $\bm{T^*\approx130}$\,K. However, no relevant anomalies are detected in the elastoresistance data near $\bm{T^*}$, which excludes the even-parity ferro-orbital nematicity often found in other superconductors. Moreover, in the temperature range between $\bm{T_{\rm CDW}}$ and $\bm{T^*}$, conical rotations of magnetic field yield a distinct first-order phase transition,  indicative of time-reversal symmetry breaking. These results provide thermodynamic evidence for the emergence of an odd-parity nematic order, implying that an exotic loop-current state precedes the CDW in CsV$_3$Sb$_5$.}

The kagome lattice with strong inherent geometrical frustration provides a rich platform for exploring various kinds of exotic phases, including quantum spin liquid, in which spins are quantum mechanically entangled, and topologically nontrivial electronic structures such as magnetic Weyl semimetals and strongly correlated flat bands.   Kagome metals $A$V$_3$Sb$_5$ ($A=$ K,\,Rb,\,Cs) possessing CDW and superconducting instabilities, whose crystal structure consists of two-dimensional (2D) V-Sb kagome layers (Fig.\,1\FigCap{a}), have aroused great interest because of their unique properties~\cite{ortiz2019new,ortiz2020cs}.  This system exhibits the CDW transition characterized by the formation of the star of David pattern or its inverse, which is considered to be driven by the nesting of saddle points near the $M$ point \cite{ortiz2021superconductivity,zhou2021origin}. Below $T_{\rm CDW}$, several intriguing phenomena such as nematicity, time-reversal symmetry breaking (TRSB), and superconductivity have been reported. The nematic phase is characterized by the appearance of a two-fold ($C_2$) symmetry, breaking the six-fold ($C_6$) symmetry of the crystal lattice. Of particular interest is the chiral loop current order or chiral flux phase, which is closely related to the TRSB~\cite{park2021electronic,tazai2022charge}.  Chiral loop-current orders have long been explored in low-dimensional electron systems, especially in the pseudogap states in cuprate superconductors~\cite{varma1997non,li2008unusual,pershoguba2013proposed}  and iridates~\cite{zhao2016evidence,jeong2017time,murayama2021bond}, but the presence of such an order remains open. 

An important question concerns the intertwining relationship between the TRSB and nematicity. Related to this, there is a fundamental question as to whether the unconventional non-phonon-mediated superconducting pairing mechanism is operative in this system~\cite{tan2021charge,tazai2022mechanism}. In $A$V$_3$Sb$_5$, however, the origins of the nematicity and TRSB have been widely debated, with contradicting results. Currently lacking is a consensus on the onset temperatures of the nematicity and TRSB, including its presence/absence, that strongly depend on the experimental methods.

 Several groups have reported that the onset temperature of nematic transition $T_{\rm nem}$ coincides with $T_{\rm CDW}$~\cite{xiang2021twofold}. 
On the other hand, it has also reported that $T_{\rm nem}$ is around 30\,K~\cite{nie2022charge,yu2021evidence}, much lower than $T_{\rm CDW}$. 
Thus, it has been controversial whether the nematicity is coupled to the CDW formation. The TRSB has also been highly controversial.  Similar to the nematic transition, different onset temperatures $T_{\rm TRSB}$ have been reported. While the anomalous Hall effect suggests $T_{\rm TRSB}=T_{\rm CDW}$~\cite{yang2020giant,yu2021concurrence}, some of the muon spin relaxation ($\mu$SR) measurements~\cite{mielke2022time} and nonreciprocal transport measurements suggest that $T_{\rm TRSB}$ is well below $T_{\rm CDW}$~\cite{guo2022switchable}. 
The polar Kerr rotation measurements are even more controversial; while the appearance of the TRSB at $T_{\rm CDW}$ has been reported~\cite{hu2022time,xu2022three}, the absence of the TRSB down to low temperatures has also been suggested~\cite{saykin2022high}. The emergence of the chiral loop currents that can lead to the TRSB has also been pointed out. 
Moreover, the chirality of these flux currents is tunable by applying magnetic fields along the $c$ axis, as suggested by the scanning tunneling microscopy (STM) studies in KV$_3$Sb$_5$~\cite{jiang2021unconventional}. In addition, nonreciprocal transport measurements report the switching of electronic magneto-chiral anisotropy~\cite{guo2022switchable}. In stark contrast, no observation of such chiral flux currents has been reported by spin-polarized STM~\cite{li2022no}. 

It has been discussed that CDW orders can be generally described by the modulation of electron hopping. In the presence of real hopping amplitudes between the neighboring bonds, the conventional CDW (rCDW) is formed. On the other hand, when the hopping amplitude is purely imaginary, loop current order that can be seen as imaginary CDW (iCDW) is formed~\cite{park2021electronic}. The TRSB in the CDW state, which may be present in CsV$_3$Sb$_5$, suggests the coexistence of rCDW and iCDW. Then, of particular interest is, in analogy with the pseudogap state above the CDW transition as debated in underdoped cuprates, whether a pure imaginary CDW state emerges at the temperature distictly above $T_{\rm CDW}$ in CsV$_3$Sb$_5$. To fully elucidate the underlying physics, it is necessary to precisely measure the signal from the symmetry-broken phases and determine the exact temperature of these phase transitions using thermodynamic bulk probes. It is also important to use probes that can distinguish even- and odd-parity orders. Here, to obtain deeper insights into the symmetry-breaking phenomena in CsV$_3$Sb$_5$, we performed extremely high sensitive measurements of the in-plane magnetic anisotropy using magnetic torque, which provides thermodynamic information of the nematic transition. We also measured the nematic susceptibility with elastoresistance, which is an even-parity probe. The combination of these measurements provides pivotal information on the order parameter.

Measurements of the magnetic torque $\boldsymbol{\tau}=\mu_0V{\bm M}\times {\bm H}$ have a high sensitivity for detecting the magnetic anisotropy, where $\mu_0$ is vacuum permeability, $V$ is the sample volume and ${\bm M}$ is the magnetization induced by external magnetic field ${\bm H}$. Torque is a thermodynamic observable that is equal to the derivative of the free energy with respect to angular displacement. We performed torque measurements by rotating ${\bm H}=H(\sin\theta\cos\phi,\sin\theta\sin\phi,\cos\theta)$ conically as a function of $\phi$, keeping $\theta$ constant, where $\theta$ is the polar angle measured from the $c$ axis and $\phi$ is the azimuthal angle measured from the $a$ axis. For the torque magnetometry, we use a highly sensitive piezoresistive cantilever. The experimental setup for this measurement is illustrated in Fig.\,1\FigCap{b}. To measure the angular variation of the magnetic torque, we use a system with two superconducting magnets generating magnetic fields in two mutually orthogonal directions and a cryostat set on a mechanical rotating stage at the top of a dewar. By computer controlling the two superconducting magnets and the rotating stage, ${\bm H}$ can be precisely rotated within $ac$ and $ab$ plane and conically around the $c$ axis. In general, the $i$ component of magnetization $M_{i}$ is given by the susceptibility tensor $\chi_{ij}$ as $M_{i}=\sum_{j}\chi_{ij}H_j$, when we take rectangular coordinates $(i,j=x,y,z)$. Hereafter we define $x$ ($z$) along the $a$ ($c$) direction and $y$ perpendicular to $a$ in the $ab$ plane. To check the sensitivity of the present setup, we measured the temperature dependence of magnetic anisotropy $\chi_{zz}-\chi_{xx}(=\chi_{cc}-\chi_{aa})$ on a very tiny single crystal (sample \#0) with the size of  90$\times$90$\times$60$\times$\,$\mu$m$^3$ at $\mu_0H=7$\,T, which is obtained from the $ac$ rotation experiments~\cite{okazaki2011rotational} (Fig.\,1\FigCap{c}). Below $T_{\rm CDW}$, $\chi_{zz}-\chi_{xx}$ decreases rapidly and increases below $\sim$60\,K, which well reproduces the previous results~\cite{PhysRevLett.129.056401}.

When ${\bm H}$ is rotated within the hexagonal $ab$ plane ($\theta=90^{\circ}$), the torque measurements present a thermodynamic test whether or not the electronic state breaks the $C_6$ crystal symmetry~\cite{okazaki2011rotational,sato2017thermodynamic,murayama2019diagonal,murayama2021bond}. In this configuration, $\tau(\phi, \theta=90^{\circ})$ in a hexagonal magnet is expected to have a six-fold oscillation term as function of $\phi$, but we find that this six-fold term is negligibly small in the present nonmagnetic metal CsV$_3$Sb$_5$. In such a case, the rotational symmetry breaking can be detected sensitively by the in-plane torque~\cite{okazaki2011rotational} 
\begin{equation}
	\tau(\phi,90^{\circ})=\frac{1}{2}\mu_0VH^2[(\chi_{xx}-\chi_{yy})\sin2\phi+2\chi_{xy}\cos2\phi].
\end{equation}

Two-fold oscillations of $\tau(\phi,90^{\circ})$ appears when the hexagonal symmetry is broken by a new electronic state: $C_6$ rotational symmetry breaking is revealed by $\chi_{xx}\neq\chi_{yy}$ or $\chi_{xy}\neq0$.

When the nematicity appears in the hexagonal lattice, the formation of nematic domains is naturally expected. In the presence of a large number of domains, the twofold oscillations due to nematicity are canceled out. Therefore, we use crystals with sizes of  120\,$\times$\,110\,$\times$\,70\,$\mu$m$^3$ (sample \#1) and 60\,$\times$\,70\,$\times$\,40\,$\mu$m$^3$ (sample \#2) in the in-plane torque measurements. Samples \#0, \#1, and \#2 are cut from the same large single crystal. If crystals are small enough to contain a small number of domains with imbalanced volumes, the twofold oscillations can be detected when the nematicity appears. Note that according to the recent optical studies, the typical length scale of the structural domain is $\sim $100\, $\mu$m~\cite{xu2022three}.
Figure\,2\FigCap{a} displays the raw data of the magnetic torque $\tau(\phi,90^{\circ})$ of Sample \#1 in magnetic field of $\mu_0H=$7\,T rotating within the $ab$ plane at several temperatures. All torque curves are perfectly reversible with respect to the field rotation direction, indicating no ferromagnetic impurities. At lower temperatures below 130\,K, $\tau(\phi,90^{\circ})$ exhibits distinct two-fold oscillations as a function of $\phi$.  We find that  $\tau(\phi,90^{\circ})$  can be fitted as,  
\begin{equation}
\tau(\phi,90^{\circ})=\tau_{2\phi}^{ab}\sin2\phi,
\end{equation}
with positive $\tau_{2\phi}^{ab}$, namely $\chi_{xx}-\chi_{yy}>0$, indicating that the nematicity is directed to the $a$ axis (inset of Fig.\,2\FigCap{b}).  Figure\,2\FigCap{b} shows the temperature dependence of $\tau_{2\phi}^{ab}$  for sample \#1 and sample \#2. At high temperatures, $\tau_{2\phi}^{ab}(T)$ is zero, which ensures that the field is controlled within the $ab$ plane with no obvious misalignment effect. As the temperature is lowered, $\tau_{2\phi}^{ab}(T)$ becomes finite exhibiting a kink at $T^*=130$\,K and grows rapidly. For sample \#1, $\tau_{2\phi}$ shows a kink at $T_{\rm CDW}$ and the slope of $\tau_{2\phi}$ tends to be reduced.  For sample \#2, $\tau_{2\phi}$ exhibits a sharp peak at $T_{\rm CDW}$.  These demonstrate that the CDW transition seriously influences the nematicity.

The present high-resolution torque measurements reveal a rotational symmetry breaking at $T^*$, providing thermodynamic evidence for the nematic transition well above $T_{\rm CDW}$. This is in contrast to previous studies, which report that the nematicity appears at or well below $T_{\rm CDW}$. As shown in Fig.\,2\FigCap{b}, in the temperature range between $T_{\rm CDW}$ and $T^*$, $\tau_{2\phi}^{ab}(T)$ increases linearly with temperature as $\tau_{2\phi}^{ab}(T)\propto(T^*-T)^{\beta}$ with exponent $\beta=1$. It should be noted that this critical exponent shows large deviations from all known results of 2D nematic transitions; for example, those of the mean-field ($\beta=1/2$) and the 2D Ising model ($\beta=1/8$). Thus the observed temperature dependence implies a nontrivial phase transition at $T^*$.  

As shown in Fig.\,2\FigCap{b}, $\tau_{2\phi}^{ab}(T)$ is suppressed below $T_{\rm CDW}$. For sample \#1, at around 80\,K, $\tau_{2\phi}^{ab}(T)$ peaks  and decreases with decreasing $T$.  However, at low temperatures,  $\tau_{2\phi}^{ab}$ remains finite and becomes nearly temperature independent below 40\,K.  It is tempting to consider that the reduction of $\tau_{2\phi}^{ab}(T)$ below 20\,K is related to the anomalies reported by the non-reciprocal transport and $\mu$SR measurements~\cite{guo2022switchable,yu2021evidence}. For sample \#2, $\tau_{2\phi}^{ab}$ rapidly decreases  below $T_{\rm CDW}$ and shows a gradual suppression  below $\sim$80\,K.  The discrepancy between sample \#1 and \#2 probably arises from the size of the nematic domain.

A key result in the in-plane torque measurements is that rotational symmetry breaking is observed above the CDW transition temperature. In metallic states in strongly correlated materials, such as iron-based superconductors, ferro-orbital nematicity that breaks the crystal rotational symmetry is often reported, where the divergent behavior of nematic susceptibility is observed~\cite{chu2012divergent}. The nematic susceptibility can be extracted by elastoresistance measurements, in which changes in resistance $R$ are measured when the anisotropic strain is introduced by using a piezoelectric actuator. As the strain can couple linearly to even-parity nematic orders, this technique can quantify the susceptibility to nematicity above the nematic transition, which is related to the fluctuations of a rotational-symmetry-breaking order of even parity. Thus, this method is complementary to the magnetic torque measurements, which sensitively detect the nematic order parameter below the transition. 

The relative change in resistance $\Delta R$ with the strain $\epsilon$ is given by the following relation using the elastoresistance tensor components $m_{ij}\ (i, j=1\sim6)$:
\begin{equation}
	(\Delta R/R)_i = \sum^6_{j = 1}m_{ij}\epsilon_j,
\end{equation}
where $i$ and $j$ follow the Voigt notation (the values of 1 to 6 correspond to $xx, yy, zz, yz, zx$, and $xy$, respectively).
The nematic susceptibility in the $D_{6h}$ point group corresponds to the elastoresistance coefficient $m_{11} - m_{12}$ with the $E_{2g}$ symmetry, while the totally symmetric mode with the $A_{1g}$ symmetry is given by the elastoresistance coefficient $m_{11} + m_{12}$ (see Fig.\,3\FigCap{a}).
We used two independent methods, four-probe and modified Montgomery methods~\cite{kuo2016ubiquitous}, to evaluate the strain-induced changes in resistance (Fig.\,3\FigCap{b},\FigCap{c}), from which we obtained the elastoresistance coefficients with the $E_{2g}$ and $A_{1g}$ irreducible representations given by
\begin{align}
	m_{11} - m_{12}=\frac{d[(\Delta R/R)_{xx}-(\Delta R/R)_{yy}]}{d[\epsilon_{xx}-\epsilon_{yy}]}&=
	\frac1{1+\nu_p}\frac{d[(\Delta R/R)_{xx}-(\Delta R/R)_{yy}]}{d\epsilon_{xx}},\\
	m_{11} + m_{12}=\frac{d[(\Delta R/R)_{xx}+(\Delta R/R)_{yy}]}{d[\epsilon_{xx}+\epsilon_{yy}]}&=
	\frac1{1-\nu_p}\frac{d[(\Delta R/R)_{xx}+(\Delta R/R)_{yy}]}{d\epsilon_{xx}},&
\end{align}
respectively, where $\nu_p = -\epsilon_{yy}/\epsilon_{xx}$ is the effective Poisson’s ratio between the $x$ and $y$ directions. Here, the strain $\epsilon$ acts as a field, and the nematic order parameter is evaluated by the induced resistance anisotropy ($\Delta R/R$). Figure\,3\FigCap{d} shows the temperature dependence of the rotational-symmetry-breaking $E_{2g}$ component, which corresponds to the parity-even nematic susceptibility, in comparison with that of isotropic $A_{1g}$ component. The two methods give consistent results showing sharp anomalies at $T_{\rm CDW}$, which is related to the resistive kink anomaly associated with the CDW transition (see Supplementary Information). Most importantly, the temperature dependence of the $E_{2g}$ component shows no discernible anomaly around $T^*$ and no divergent behavior above $T^*$. The observed absence of the divergent behavior indicates that the nematic order below $T^*$ found in CsV$_3$Sb$_5$ is essentially different from the even-parity ferro-orbital order as found in iron-based superconductors.  

In order to gain further insight into the nematic phase transition, we investigated the influence of the $c$-axis component of the field. For this purpose, we rotated ${\bm H}$ ($\mu_0|{\bm H}|=3$\,T) conically as a function of $\phi$ keeping $\theta=60^{\circ}$ in the torque magnetometry measurements for sample \#1 (Fig.\,4\FigCap{a}). At 120\,K, $\tau(\phi,60^{\circ})$ shows two-fold oscillation, similar to  $\tau(\phi,90^{\circ})$.   Surprisingly, $\tau(\phi,60^{\circ})$ below $\sim 120$\,K exhibits clear jumps as well as hysteresis loops as a function of $\phi$ near $\phi \sim -30^{\circ}$ and $\sim 150^{\circ}$, revealing the occurrence of a first-order phase transition. What is remarkable is that $\tau(\phi,60^{\circ})$ exhibits a local minimum at $\phi\sim45^{\circ}$ at 112\,K, while it shows a maximum at 120\,K, indicating that the phase of $\tau(\phi,60^{\circ})$ between $\phi\sim -30^{\circ}$ and $150^{\circ}$ below $\sim120$\,K shifts around 120$^{\circ}$-180$^{\circ}$ from that at higher temperatures.  In Fig.\,4\FigCap{b}, $\bm{H}$ is conically rotated at different $\theta$ angles, keeping in-plane field constant ( $\mu_0H_{ab}=2.1$\,T) at 100\,K. The jump suddenly appears when $\theta$ exceeds $30^{\circ}$, indicating the presence of the threshold field along the $c$ axis.  
Figure\,4\FigCap{c} depicts the magnitude of the torque jump in the angular variation $\Delta \tau$ ($>$0 at $\sim-30^{\circ}$) obtained from Fig.\,4\FigCap{a}. The jump appears slightly below $T^*$ and its magnitude is rapidly enhanced with decreasing temperature. Then the jump height is significantly suppressed by the CDW transition. The direction of the jump becomes opposite ($\Delta \tau<0$) below $\sim85$\,K and disappears  below $\sim70$\,K.  
 
Since CsV$_3$Sb$_5$ is a paramagnetic system with no local magnetic moments,  the angular variation should be a continuous sinusoidal curve, as observed in the in-plane field rotation of $\tau_{2\phi}^{ab}$.    The jump as well as hysteresis loop in the angular variation of torque are typically observed in ferromagnetic and ferrimagnetic materials.   However,  we can exclude the possibility that the jump and hysteresis loop are caused by the ferromagnetic impurities. In fact, no jump and hysteresis is observed in the in-plane field rotation. In addition, the jump disappears at low temperatures below $T_{\rm CDW}$, demonstrating that the jump is closely related to the CDW state.  Moreover, we paid careful attention during the crystal growth to prevent the magnetic impurity contamination by using non-magnetic crucibles.  The jump of the magnetic torque with rotating $\phi$ is observed in sample \#2, when the finite  $c$-axis component of ${\bm H}$ is applied, confirming the first-order transition induced by the $c$ axis field, as shown in SI.  Therefore, the present results demonstrate the emergence of an anomalous paramagnetic state below $T^*$, whose magnetic response bears striking resemblance with the ferromagnetic state.  The present results imply that the nonmagnetic electronic nematic order directly couples with the external magnetic field.  The sensitivity to magnetic fields of only a few Tesla indicates a linear coupling with the magnetic field. It should be noted that nematic order does not linearly couple with the magnetic field  for the even-parity case.  Therefore, the present results provide evidence for the emergence of an odd-parity nematic order, i.e. the TRSB below $T^*$.

A plausible explanation is that the hysteresis loop and jump are caused by the nematic domains. The domain structure is expected to be determined by the complicated inter-domain coupling and the pinning energy of domains. Moreover, the domains also couple with the CDW order. These make the temperature and angular dependence of the hysteresis loop complicated. It is interesting to note  that the presence of the chiral domains have been recently reported by Kerr rotation measurements~\cite{xu2022three}, although its relationship with the present results is unknown.
 
The present results demonstrate a tantalizing relationship between the nematicity and TRSB, both of which appear around $T^*$ well above $T_{\rm CDW}$. It is natural to consider that the nematicity is caused by the current order.  Given that the system is paramagnetic, the TRSB can be attributed to the charge current order, i.e. iCDW. We note that the current order can also lead to the nematicity. We propose two possibilities to explain the first-order transition observed in the conical rotation of ${\bm H}$; single- and triple-${\bm q}$ loop current scenarios. In the single-${\bm q}$ scenario, 1D currents flow along the bond direction and adjacent currents flow alternately in opposite directions, which gives rise to the rotational symmetry breaking of the 2D kagome layer, as illustrated in Fig.\,4\FigCap{d}.  The inset illustrates a schematic nematic Fermi surface realized by the single-${\bm q}$  iCDW. When the Fermi surface is nematic, the van-Vleck susceptibility leads to the in-plane
anisotropy of magnetic susceptibility, which is observable by the magnetic torque measurement ~\cite{kawaguchi2020spin}.  Unfortunately, however, the exact phase shift of $\tau (\phi, \theta)$ across the jumps is difficult to determine because of the resolution and large hysteresis loops associated with jumps, although the phase shift is between 120$^{\circ}$ to 180$^{\circ}$.  If the phase shift is 120$^{\circ}$, the first-order phase transition may originate from the transition between the single-${\bm q}$ phases with different ${\bm q}$ orientations. Also, the first-order transition may be caused by the transition from the single-$\bm{q}$ to double-$\bm{q}$ state. This picture is consistent with the 180$^{\circ}$ periodicity with respect to $\phi$ of the first order transition in the conical rotation. 
In the triple-${\bm q}$ case, the first order transition is caused by the flipping of loop current by the $c$ axis component of $\bm{H}$ (Fig.\,4\FigCap{e}). In this scenario, the 2D kagome layer itself does not break the rotational symmetry, but the stacking of kagome layer induces the nematicity, similar to the nematicity proposed in the 2$a_0$ $\times$2$a_0$ $\times$2$c_0$ CDW state. Further study is required to clarify the origin of the odd-parity nematic order and the first-order phase transition.  According to recent theoretical studies, chiral triple-$\bm{q}$ iCDW component induced by applying finite magnetic fields to the single-$\bm{q}$ iCDW state.  It is noteworthy that triple-$\bm{q}$ iCDW linearly couples to the magnetic field, giving  rise to the first-order transition.

Above we report the emergence of nematicity and first-order transition below $T^*$. However, no clear anomalies around $T^*$ have been reported by other experiments so far. Very recent work on X-ray diffraction shows a reduction of the $a$-axis lattice parameter below $\sim 130-160$\,K~\cite{PhysRevLett.129.056401}, although other X-ray measurements show no such a behavior, which deserves further studies using different techniques~\cite{preprint}.  For the TRSB, we observe the first-order transition only when a finite $c$-axis field is applied. This suggests that the internal magnetic field associated by TRSB is strongly enhanced by the $c$-axis field. Then the TRSB may occur even at zero field but high resolution measurements are required for the detection. In tetragonal systems, it has been proposed that the $c$-axis component of the magnetic field can induce monoclinic distortion, enabling the magnetic field to couple to the nematic order parameter when the order parameter is oriented in-plane~\cite{varma2020thermal}. Therefore, exploring if a similar mechanism can arise in hexagonal systems is intriguing. Finally, while a few Kerr measurements report different results about the TRSB below $T_{\rm CDW}$, none of them has reported the TRSB above $T_{\rm CDW}$.  If the system is in the single-${\bm q}$ phase, the inconsistency between these and our results may be explained because the time-reversal symmetry is preserved macroscopically, and no Kerr effect is expected. 

Electronic nematic states that break rotational symmetry of underlying lattice have aroused great interest in various condensed-matter systems, including superconductors, quantum magnets, and topological materials. Until now, almost all nematic states have even parity and are uniform (${\bm q}=0$). The quest for an odd-parity electronic nematic state has been extremely challenging due to the lack of appropriate candidate materials. We showed that the present kagome metal CsV$_3$Sb$_5$ is a prime candidate for such a state. In particular, if the system is in the single-${\bm q}$ phase, the translational symmetry is also broken, known as the smectic state. Combined with the TRSB, CsV$_3$Sb$_5$ provides a fertile playground to investigate the exotic phases with broken multiple symmetries.

\noindent
\section*{Methods}
\noindent
\textbf{Sample growth}\newline\noindent
High-quality single crystal samples of CsV$_3$Sb$_5$ were grown by the self-flux method. 

\noindent		
Samples for elastoresistivity measurements were grown at the University of California, Santa Barbara. Elemental reagents of Cs (Alfa, 99.98\%), V (Sigma, 99.9\%), and Sb (Alfa, 99.999\%) were mixed using mechanochemical methods to produce a precursor powder with a composition of approximately Cs$_{20}$V$_{15}$Sb$_{120}$. The subsequent powder was loaded into 2\,mL high-density alumina (Coorstek) crucibles and loaded into steel tubes under 1 atm of argon.  Samples were heated to 1000\,$^{\circ}$C at 200\,$^{\circ}$C/hr and subsequently soaked at 1000\,$^{\circ}$C for 12\,h. Samples were then cooled relatively quickly to 900\,$^{\circ}$C at 5\,$^{\circ}$C/hr and then slow cooled to 600\,$^{\circ}$C at 1\,$^{\circ}$C/hr. Fluxes were cooled completely and crystals were extracted mechanically. The resulting crystals are thin hexagonal flakes with a metallic silver luster. 

\noindent
Samples for torque magnetometry measurements were grown at Nanjing University. To exclude the possibility of magnetic impurity inclusion, we used non-magnetic alumina crucibles. A more detailed description can be found in Ref.~\cite{xiang2021twofold}. 
\\

\noindent
\textbf{Torque measurements}\newline\noindent
Magnetic torque was measured by the piezoresistive micro-cantilever technique.
Tiny single crystals of CsV$_3$Sb$_5$ were carefully mounted onto the cantilever using a tiny amount of glue. The piezoresistive responses were measured by the electrical bridge circuit. The magnetic torque $\tau$ was measured as $\tau = \frac{at^2}{2\pi_L} \frac{\Delta R}{R_s}$, where $a$ is the leg width, $t$ is the leg thickness, $\pi_L$ is the piezoresistive coefficient, and $R_s$ is the resistance of the cantilever. To precisely measure the in-plane magnetic torque and apply conical magnetic fields, we used a 2D vector magnet with a mechanical rotator. \\ 

\noindent
\textbf{Elastoresistance measurements}\newline\noindent
The elastoresistance measurements were carried out by applying an in-plane uniaxial strain to the CsV$_3$Sb$_5$ single crystals using a piezoelectric device (Piezomechanik PSt150/3.5$\times$3.5/7). After fixing the samples on the piezo stack with glue (ARALDITE RT30), they were cleaved to a thickness of several tens of $\mu$m so that the strain provided by the piezo stack could be transferred to the top surfaces of the samples where the electrical contacts were made after cleavage.
We measured the elastoresistance in two different configurations (four-probe and modified Montgomery methods, for more details, see Supplementary Information), in which the amount of the strain $\epsilon_{xx}$ along the $x$ direction was controlled by applying the voltage to the piezo stack and monitored by the strain gauge attached on the other side, while the orthogonal strain $\epsilon_{yy}$ was calculated by Poisson’s ratio of the piezo stack calibrated beforehand.\\

\noindent
{\bf ACKNOWLEDGMENT:} 
 A part of this work was supported by CREST (No. JPMJCR19T5;  Y.M. and T.S.) from Japan Science and Technology (JST), Grants-in-Aid for Scientific Research (KAKENHI)  (Nos. 18H05227, 18H03680, 18H01180,21K13881)  and Grant-in-Aid for Scientific Research on innovative areas ‘Quantum Liquid Crystals’ (No. JP19H05824) from Japan Society for the Promotion of Science (JSPS) (T.S.). S.D.W. and B.R.O. gratefully acknowledge support via the UC Santa Barbara NSF Quantum Foundry funded via the Q-AMASE-i program under award DMR-1906325.\\
{\bf Data Availability Statement:}
The data that support the findings of this study are available from the corresponding authors upon reasonable request.\\
{\bf Author contributions:} 
T.A., T.S. and Y.M. conceived and supervised the study.  T.A.,  T.K., K.O., S.S. and T.G. performed torque measurements. A.O., Y.Kag. and K.H. performed elastoresistivity measurements. B.R.O., S.D.W., Q.Li and H.-H.W. synthesized single crystals.  T.A., A.O, Y.Kag., T.K., K.O. and K.H. analyzed the data with inputs from S.S., Y.Koh., Y.Kas. and H.M. Theoretical models were provided by R.T. and H.K. All authors discussed the results and contributed to writing the manuscript.

\noindent


\newpage 
\begin{figure}[t]
    \label{fig:basic}
	\includegraphics[clip,width=\linewidth]{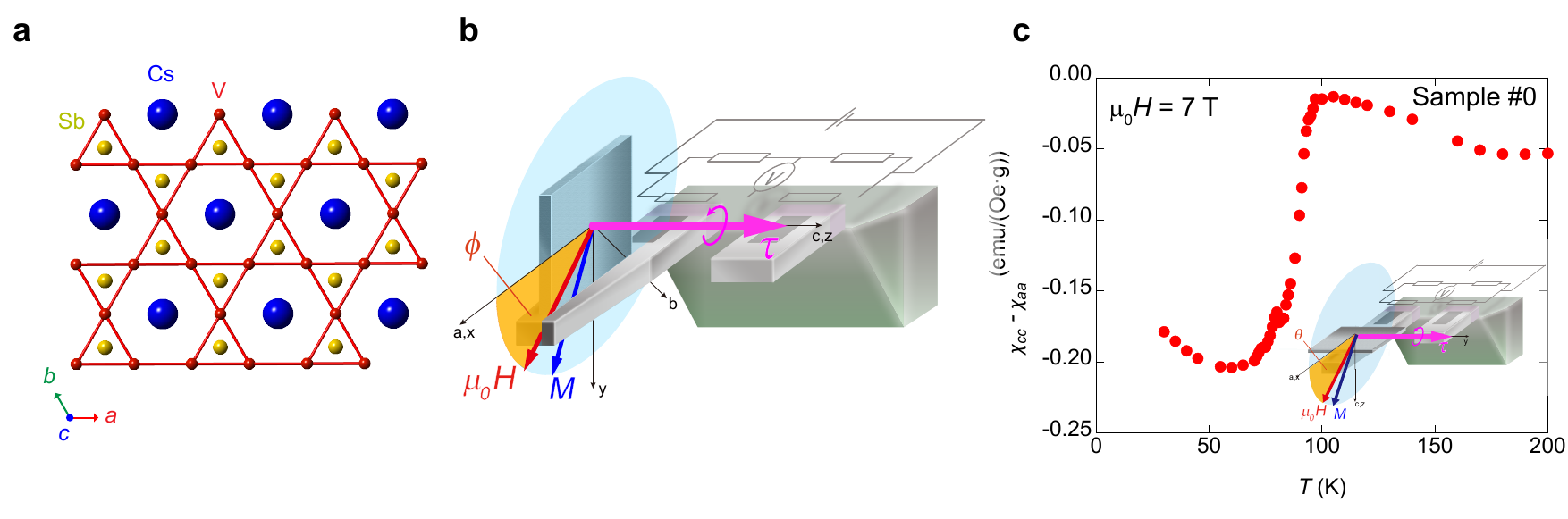}
	\caption{\textbf{Crystal structure, experimental setup, and magnetic anisotropy.} \FigCap{a}, The crystal structure of CsV$_3$Sb$_5$. Red solid lines indicate the kagome layer of V-atoms. \FigCap{b},  Experimental setup of in-plane magnetic torque measurements. \FigCap{c}, Temperature dependence of the magnetic susceptibility anisotropy $\Delta\chi$=$\chi_{zz}$-$\chi_{xx}$ of sample \#1. The inset depicts the setup of out-of-plane  torque measurements.
	}
\end{figure}

\begin{figure}[h]
    \label{fig:nematic}
	\includegraphics[clip,width=1.1\linewidth]{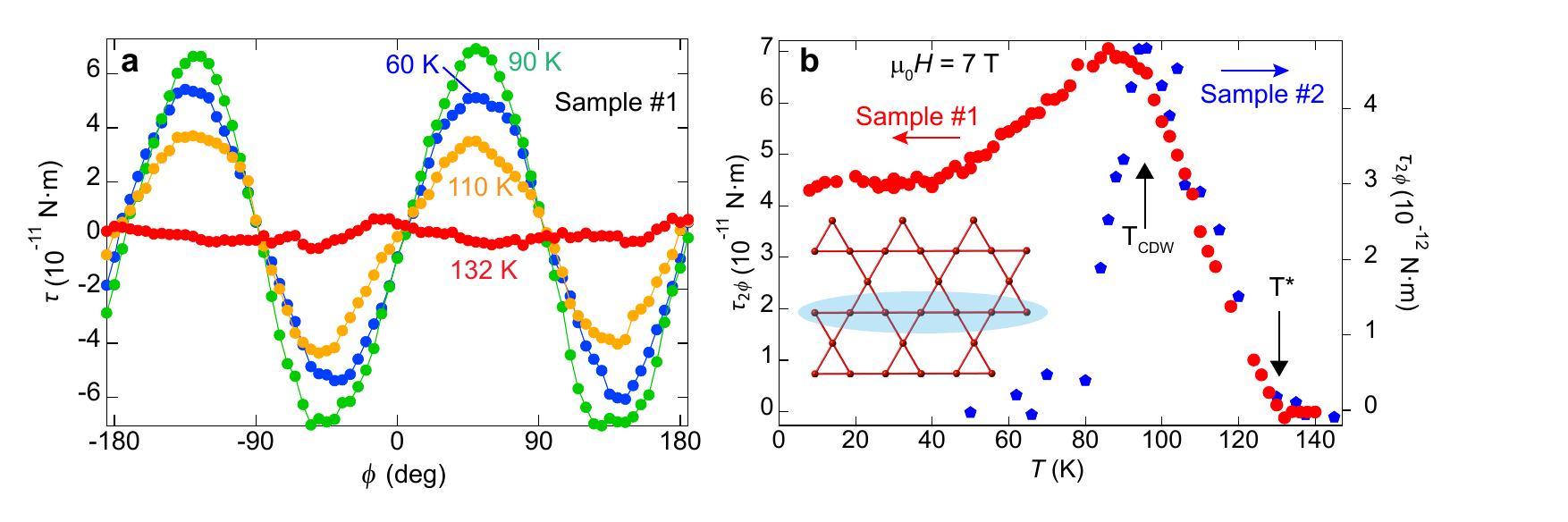}
	\caption{\textbf{In-plane variation of magnetic torque.} \FigCap{a}, The magnetic torque $\tau$($\phi$) in the in-plane magnetic field of $\mu_0H=$7\,T as a function of the azimuthal angle $\phi$. \FigCap{b}, Temperature dependence of the amplitude of twofold oscillations $\tau_{2\phi}$. To obtain $\tau_{2\phi}$, the angular-dependent curves are fitted by $\sin 2\phi$. The inset shows the direction of the nematicity (light blue ellipse).
	}
\end{figure}

\begin{figure}[t]
    \label{fig:ER}
	\includegraphics[clip,width=\linewidth]{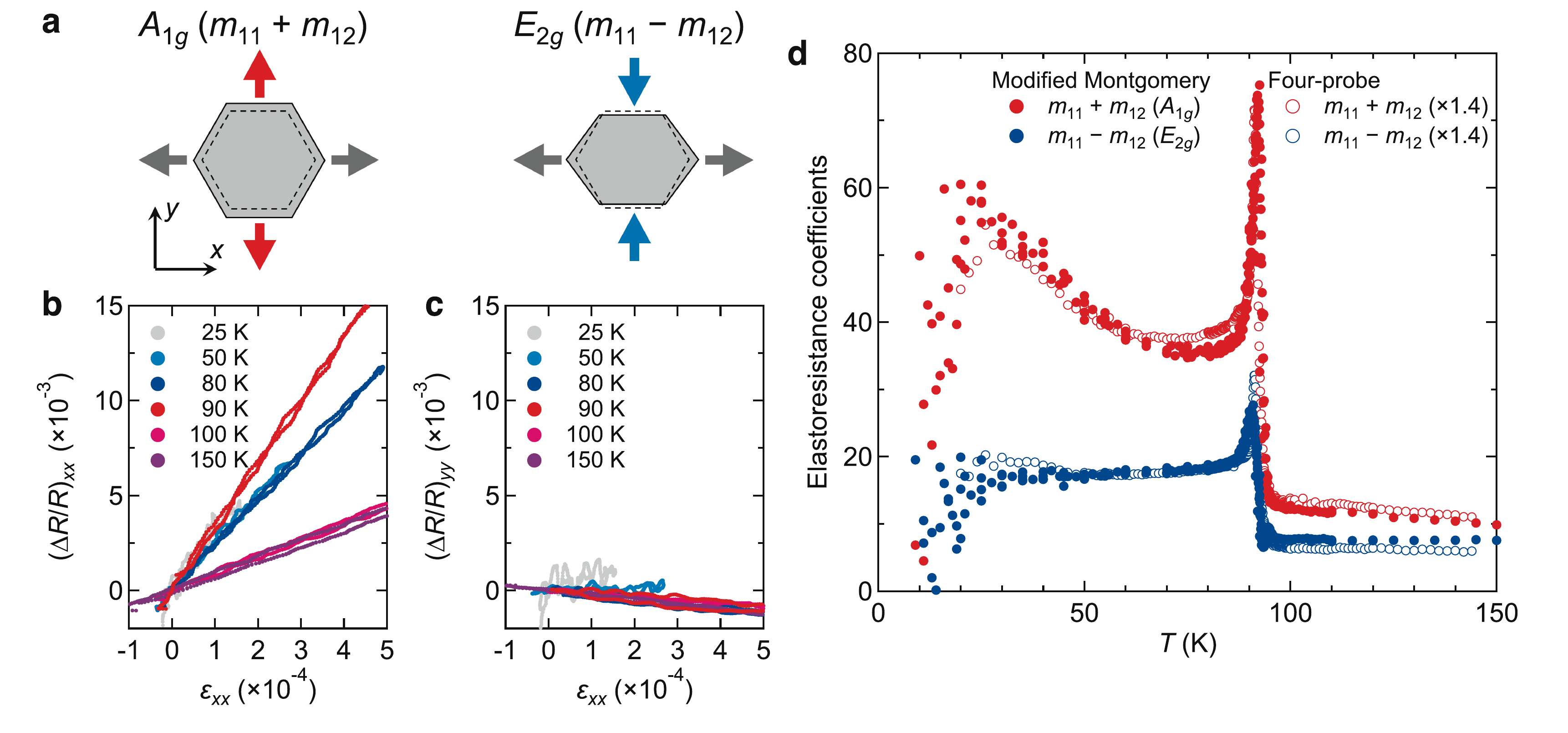}
	\caption{\textbf{Elastoresistance and nematic susceptibility.}	
		\FigCap{a}, Schematics of the strain deformations with the $A_{1g}$ (left) and $E_{2g}$ (right) irreducible representations in the hexagonal $D_{6h}$ point group.
		\FigCap{b},\FigCap{c}, Relative changes in the longitudinal resistance $R_{xx}$ (\FigCap{b}) and transverse resistance $R_{yy}$ (\FigCap{c}) as a function of strain measured at various temperatures.
		\FigCap{d}, Temperature dependence of the elastoresistance coefficients with the $A_{1g}$ (red) and $E_{2g}$ (blue) symmetries. The results obtained from the four-probe (open circle) and modified Montgomery (filled circle) methods are plotted together. The data for the four-probe method is multiplied by 1.4.
	}
\end{figure}

\begin{figure}[t]
    \label{fig:TRSB}
    \includegraphics[clip,width=\linewidth]{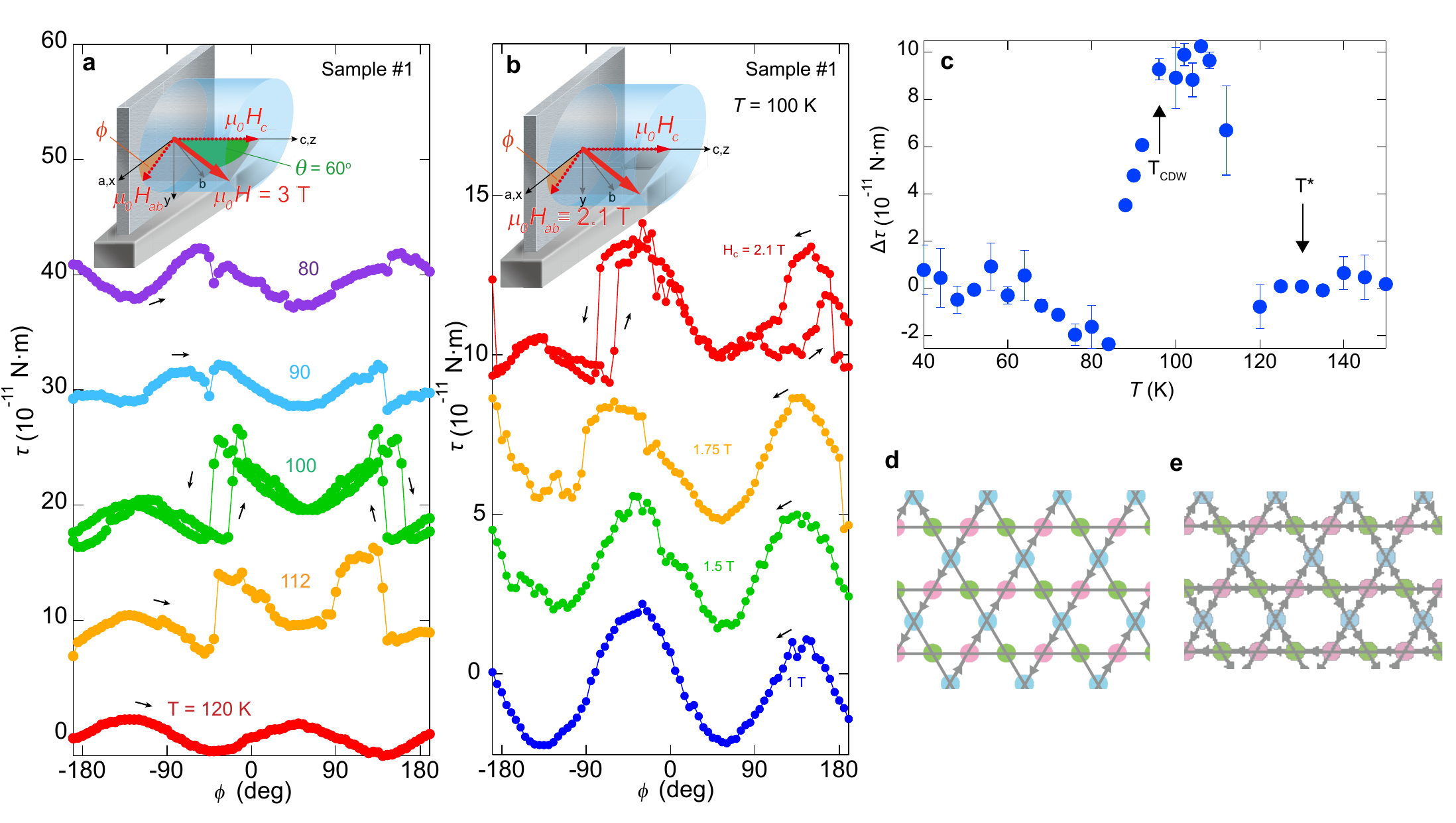}
    \caption{\textbf{Broken time-reversal symmetry and loop current orders.} \FigCap{a}, The magnetic torque $\tau$($\phi$) of sample \#1 in the conically rotated magnetic field of $|\mu_0H|=$3\,T and $\theta=$60$^{\circ}$ as a function of the azimuthal angle $\phi$. Arrows indicate the sweep directions of the field. Jumps and hysteresis loops are observed at around $\phi=$ -45$^{\circ}$ and 135$^{\circ}$. The curves are offset for clarity. \FigCap{b}, The magnetic torque $\tau$($\phi$) of sample \#1 in the conically rotated magnetic field with a fixed in-plane component $\mu_0H_{ab}=$2.1\,T. The out-of-plane component of the field $\mu_0H_{c}=$2.1, 1.75, 1.5 and 1\,T. The curves are offset for clarity. \FigCap{c}, Temperature dependence of the amplitude of the jump. Error bars are uncertainty between up and down sweeps. \FigCap{d}, Single-{$\bm q$} iCDW state. \FigCap{e}, Triple-{$\bm q$} iCDW state with chirality.
    }
\end{figure}

\end{document}


	
	\title{
Supplementary Information: Evidence for an odd-parity nematic phase above the charge density wave transition in kagome metal CsV$_3$Sb$_5$}

\author{T.\,Asaba$^{1,\dagger}$}
\author{A.\,Onishi$^{2,\dagger}$}
\author{Y.\,Kageyama$^{2,\dagger}$}
\author{T.\,Kiyosue$^1$}
\author{K.\,Ohtsuka$^1$}
\author{S.\,Suetsugu$^1$}
\author{Y.\,Kohsaka$^1$}
\author{T.\,Gaggl$^1$}
\author{Y.\,Kasahara$^1$} 
\author{H.\,Murayama$^{1,3}$}
\author{K.\,Hashimoto$^2$} 
\author{R.\,Tazai$^{4,5}$}
\author{H.\,Kontani$^4$}
\author{B.\,R.\,Ortiz$^6$}
\author{S.\,D.\,Wilson$^6$}
\author{Q.\,Li$^7$}
\author{H.-H.\,Wen$^7$}
\author{T.\,Shibauchi$^{2,*}$}
\author{Y.\,Matsuda$^{1,*}$}
	
\affiliation{$^1$Department of Physics, Kyoto University, Kyoto 606-8502 Japan}
\affiliation{$^2$Department of Advanced Materials Science, University of Tokyo, Chiba 277-8561, Japan} 
\affiliation{$^3$RIKEN Center for Emergent Matter Science, Wako, Saitama 351-0198, Japan}
\affiliation{$^4$Department of Physics, Nagoya University, Furo-cho, Nagoya 464-8602, Japan}
\affiliation{$^5$Yukawa Institute for Theoretical Physics, Kyoto University, Kyoto 606-8502, Japan}
\affiliation{$^6$Materials Department, University of California Santa Barbara, Santa Barbara, California 93106, USA}
\affiliation{$^7$National Laboratory of Solid State Microstructures and Department of Physics, Nanjing University, Nanjing 210093, China} 
	\date{\today}
	

	\maketitle

\renewcommand{\thefigure}{S\arabic{figure}}
\setcounter{figure}{0}

\section*{Elastoresistance measurements}

We performed elastoresistance measurements to examine the nematic property of \CVS{}.
In strongly correlated materials with an electronic nematic instability, the strain tensor component $\epsilon_{ij}\equiv(\partial u_i/\partial r_j+\partial u_j/\partial r_i)/2$ (where $r_{i,j}$ and $u_{i,j}$ are the $i,j$-component ($i,j=x, y, z$) of the position vector and the displacement field, respectively) acts as a conjugate field to the electronic nematic order parameter $\phi$ through the electron-lattice coupling.
Therefore, the thermodynamic nematic susceptibility $\chi_{\rm nem} \equiv \partial\phi/\partial\epsilon$ shows Curie-Weiss-like divergent behavior approaching the electronic-driven nematic order above the transition temperature. Because the normalized relative change in resistance $\Delta R/R$ is related to the strain $\epsilon$ through the elastroresistance tensor components $m_{ij}$ (see Eq.\,(3) in the main text), one can estimate $\chi_{\rm nem}$ represented by a linear combination of $m_{ij}$ by measuring the change in resistance $\Delta R/R$ as a function of $\epsilon$.

\begin{figure}[b]
	\label{fig:SI_ER}
	\includegraphics[width=0.65\linewidth]{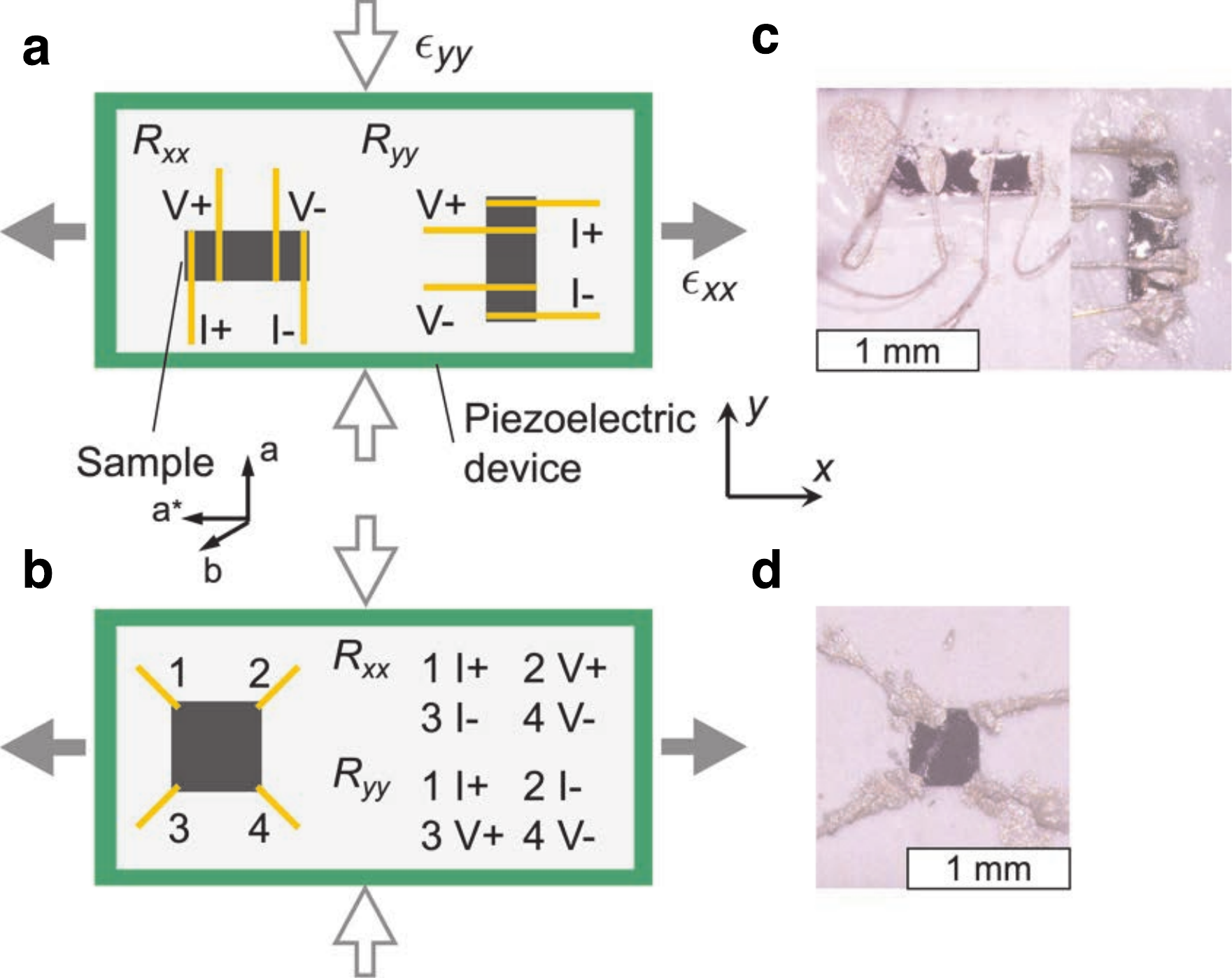}
	\caption{\textbf{Two setups for elastoresistance measurements.} {\bf a,b,} Schematic illustrations of the four-probe ({\bf a}) and modified Montgomery ({\bf b}) methods for the elastroresistance measurements. {\bf c,d,} Pictures of the four-probe ({\bf c}) and modified Montgomery ({\bf d}) methods.
	}
\end{figure}

We used two different approaches to measure the elastoresistance along the $x$ and $y$ directions;
the conventional four-probe method (Fig.\,S1\FigCap{a},\FigCap{c}) and the modified Montgomery method (Fig.\,S1\FigCap{b},\FigCap{d})~\cite{dos2011procedure,ishida2020novel}.
In the four-probe method, from a single-crystal sample, we prepared two rectangular-shaped samples (with a lateral size of $\sim \SI{1}{mm}\times\SI{0.3}{mm}$) whose short sides were parallel to the $a$- and $a^{\ast}$-axis directions, respectively (here, $\bm{a} \perp \bm{a^{\ast}}$). We set the two samples so that the $a^{\ast}$-axis direction of the samples was parallel to the strain direction of the piezo stack (see Fig.\,S1\FigCap{a},\FigCap{c}). 
In the modified Montgomery method \cite{dos2011procedure}, we prepared a square-shaped sample (with a lateral size of $\sim \SI{0.5}{mm}\times\SI{0.5}{mm}$). By switching the current-source and voltage-read probes, we obtained the elastroresistance along the $x$ and $y$ directions in the same experimental run (see Fig.\,S1\FigCap{b},\FigCap{d}).

\begin{figure}[b]
	\label{fig:SI_2}
	\includegraphics[width=1\linewidth]{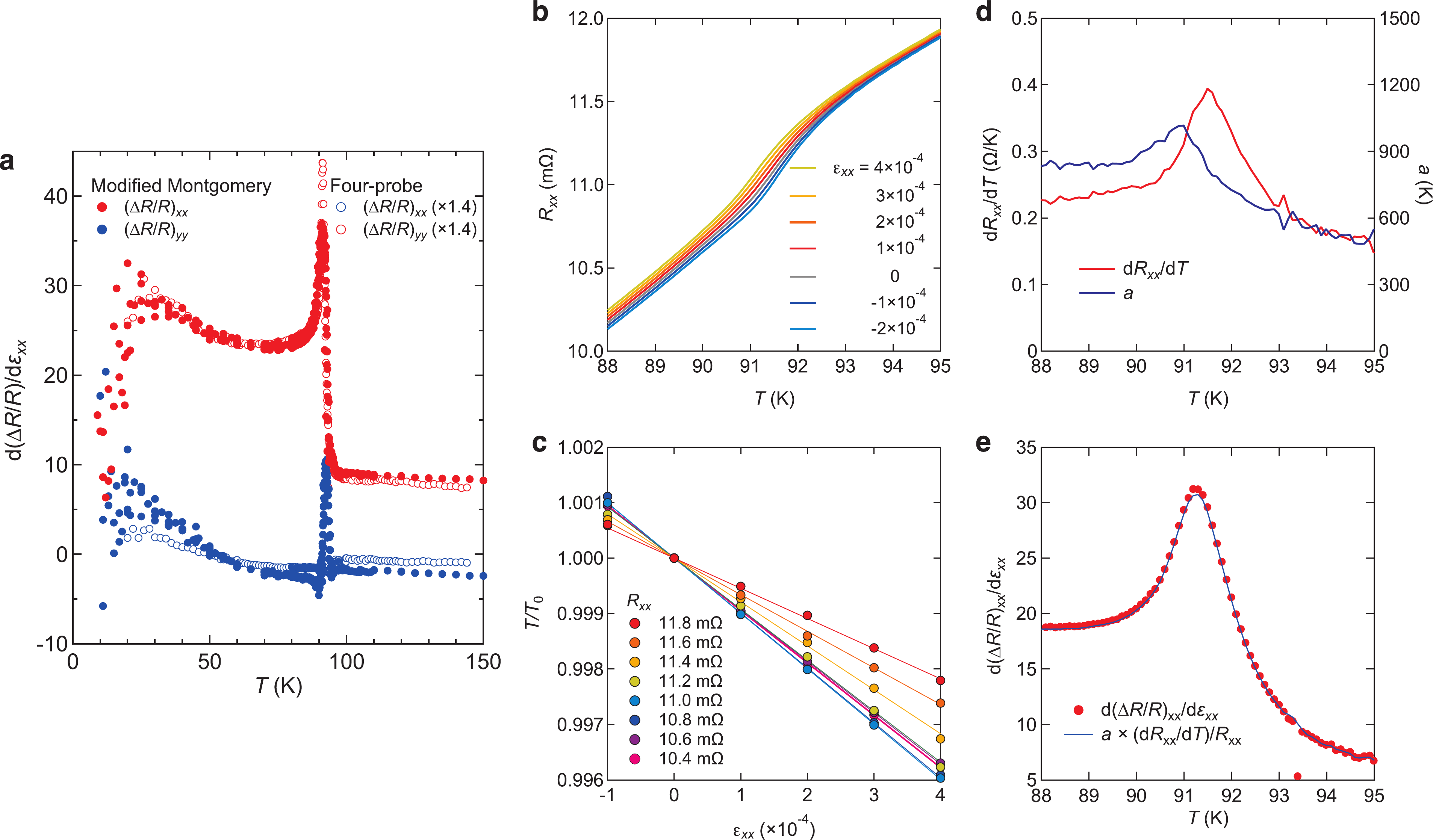}
	\caption{\textbf{Peak structure of elastoresistance coefficients at $\bm{T_{\rm CDW}}$.} {\bf a,} Temperature dependence of ${\rm d}(\Delta R/R)_{xx,yy}/{\rm d}\epsilon_{xx}$ measured by the four-probe and modified Montgomery methods. The data of the four-probe method are multiplied by 1.4. {\bf b,} Temperature dependence of $R_{xx}$ at various $\epsilon_{xx}$ values extracted from the derivative ${\rm d}(\Delta R/R)_{xx}/{\rm d}\epsilon_{xx}$. {\bf c,} Relationship between $T/T_0$ and $\epsilon_{xx}$ for a given $R_{xx}$ value. $T_0$ is the temperature at which $R_{xx}(T_0, \epsilon_{xx}=0)$ holds. {\bf d}, Temperature dependence of ${\rm d}R_{xx}(T,\epsilon_{xx}=0)/{\rm d}T$ (left) and $a(T)$ (right). {\bf e,} Temperature dependence of ${\rm d}(\Delta R/R)_{xx,yy}/{\rm d}\epsilon_{xx}$ (blue line) extracted by multiplying $({\rm d}R_{xx}(T,\epsilon_{xx}=0)/{\rm d}T)/R_{xx}$ and $a(T)$, together with the results obtained from the four-probe method (red circle).
	}
\end{figure}

Figure\,S2\FigCap{a} shows the temperature dependence of ${\rm d}(\Delta R/R)_{xx,yy}/{\rm d}\epsilon_{xx}$ extracted from the slope of $(\Delta R/R)_{xx,yy}(\epsilon_{xx})$ (see Fig.\,3\FigCap{b},\FigCap{c} in the main text). A clear peak structure has been observed at $T_{\rm CDW}$ in both directions. 
To clarify the origin of this peak, we examined how $T$ and $\epsilon_{xx}$ relate to each other under the condition that $R_{xx}(T,\epsilon_{xx})=\rm{const.}$, where the following equation is satisfied:
\begin{equation}
	{\rm d} R_{xx}(T, \epsilon_{xx}) = \frac{\partial R_{xx}(T,\epsilon_{xx})}{\partial T}{\rm d} T + \frac{\partial R_{xx}(T,\epsilon_{xx})}{\partial \epsilon_{xx}}{\rm d} \epsilon_{xx} = 0
\end{equation}
From Eq.\,S(1), we obtain the following relationship at $\epsilon_{xx}=0$:
\begin{equation}
	\left.\frac{\partial R_{xx}(T, \epsilon_{xx})}{\partial \epsilon_{xx}}\right|_{\epsilon_{xx} = 0} = - \left. \frac{\partial R_{xx}(T,\epsilon_{xx})}{\partial T}\right|_{\epsilon_{xx} = 0} \times \left.\frac{{\rm d} T}{{\rm d} \epsilon_{xx}}\right|_{\epsilon_{xx} = 0}
\end{equation}
When $T$ is constant, $\left.{\rm d}R_{xx}(T,\epsilon_{xx})/{\rm d}\epsilon_{xx}\right|_{\epsilon_{xx} = 0}$ corresponds to the left side of Eq.\,(S2). The first term of the right side, $\left. \partial R_{xx}(T,\epsilon_{xx})/\partial T \right|_{\epsilon_{xx} = 0}$, can be obtained from the temperature derivative of $R_{xx}(T)$ at $\epsilon_{xx}=0$ (see Fig.\,S2\FigCap{d}). Therefore, once one obtains information on the second term $dT/d\epsilon_{xx}$, one can derive $\left.{\rm d}R_{xx}(T,\epsilon_{xx})/{\rm d}\epsilon_{xx}\right|_{\epsilon_{xx} = 0}$. To this end, we first extracted $R_{xx}(T)$ for various $\epsilon_{xx}$ values using the derivative ${\rm d}R_{xx}(T, \epsilon_{xx})/{\rm d}\epsilon_{xx}$, as shown in Fig.\,S2\FigCap{b}. Then, by determining the crossing points $(T,\epsilon_{xx})$ between $R_{xx}(T,\epsilon_{xx})$ and $R_{xx}=\rm{const.}$, we obtained the dataset of $(T, \epsilon_{xx})$ for a constant $R_{xx}$ value (see Fig.\,S2\FigCap{c}). It can be clearly seen that a linear relationship $T=T_0-a(T)\epsilon_{xx}$ is satisfied, where $T_0$ is the temperature at which $R_{xx}(T_0, 0)$ holds and $a(T)=-\left.\frac{{\rm d} T}{{\rm d} \epsilon_{xx}}\right|_{\epsilon_{xx} = 0}$. We extracted $a(T)$ from the linear relationship between $T$ and $\epsilon_{xx}$ (see Fig.\,S2\FigCap{d}). Finally, we obtained $\left.{\rm d}R_{xx}(T,\epsilon_{xx})/{\rm d}\epsilon_{xx}\right|_{\epsilon_{xx} = 0}$ by multiplying ${\rm d}R_{xx}(T,\epsilon_{xx}=0)/{\rm d}T$ and $a(T)$, which shows a good agreement with ${\rm d}R_{xx}(T,\epsilon_{xx})/{\rm d}\epsilon_{xx}$ originally obtained in the experiment. The present experimental fact indicates that the peak structure of ${\rm d}R_{xx}(T, \epsilon_{xx})/{\rm d}\epsilon_{xx}$ comes from a combination of the peak structures of ${\rm d}R_{xx}(T,\epsilon_{xx}=0)/{\rm d}T$ and $a(T)$ at $T_{\rm CDW}$.
Such a sharp peak has not been observed in the previous elastoresistance measurements \cite{nie2022charge}.
This might be due to inhomogeneous strain over the samples caused by the large thickness of the samples or the gluing condition.

\section*{Torque magnetometry}
\noindent
{\bf In-plane and conical rotation}

Firstly, we consider a paramagnetic case, in which ${\bm M} = \chi{\bm H}$. Here, ${\bm M}$, $\chi$ and ${\bm H}$ denote magnetization, magnetic susceptibility tensor and the applied magnetic field, respectively. In this study, we consider two crystal structures: hexagonal (high-temperature phase of CsV$_3$Sb$_5$) and orthorhombic (nematic phase). In the orthorhombic systems, the magnetic susceptibility tensor can be diagonalized and given as

\begin{equation}\label{ort}
	\chi_{\rm ort}=
	\begin{pmatrix}
		\chi_{xx} & 0 & 0\\
		0 & \chi_{yy}& 0\\
		0 & 0 & \chi_{zz}\\
	\end{pmatrix}.
\end{equation}\\

Here, as shown in Fig.\,1\FigCap{b}, we define $x$ ($z$) along the $a$ ($c$) direction and $y$ perpendicular to $a$ in the $ab$ plane. In this configuration, magnetic torque $\tau(\theta,\phi)$ is given by
\begin{align}
	\tau(\theta,\phi) &= \mu_0 V {\bm M} \times {\bm H} \\
	&=\mu_0 V
	\begin{pmatrix}
		\chi_{xx} & 0 & 0\\
		0 & \chi_{yy}& 0\\
		0 & 0 & \chi_{zz}\\
	\end{pmatrix}
	\begin{pmatrix}
		H\sin\theta\cos\phi\\
		H\sin\theta\sin\phi\\
		H\cos\theta
	\end{pmatrix}
	\times
	\begin{pmatrix}
		H\sin\theta\cos\phi\\
		H\sin\theta\sin\phi\\
		H\cos\theta
	\end{pmatrix}\\
	&=\frac{1}{2}\mu_0 VH^2
	\begin{pmatrix}
		\sin 2\theta \sin \phi (\chi_{yy}-\chi_{zz}) \\
		\sin 2\theta \cos \phi (\chi_{zz}-\chi_{xx}) \\
		\sin^2\theta \sin 2\phi(\chi_{xx}-\chi_{yy}) \\
	\end{pmatrix}
\end{align}\\
Here, $\mu_0$ is vacuum permeability, $V$ is the sample volume, $\theta$ is the polar angle measured from the $c$ axis and $\phi$ is the azimuthal angle measured from the $a$ axis. In the present setup, the cantilever can only bend around $z$-axis, so this method is only sensitive to $\tau_z$. Therefore, $\tau(\theta,\phi)=\frac{1}{2}\mu_0 VH^2\sin^2\theta \sin 2\phi(\chi_{xx}-\chi_{yy})$. This indicates that $\tau(\theta,\phi)$ is only sensitive to the in-plane component of ${\bm H}$.\\

\noindent
{\bf Nematic order}

According to Neumann's principle, any of the physical properties must be invariant with respect to the symmetry elements of the crystal. In hexagonal systems, $C_6\chi C_6^{-1}=\chi$, resulting in $\chi_{xx}=\chi_{yy}$ and $\chi_{xy}=\chi_{yx}=0$. The magnetic susceptibility tensor in hexagonal systems is given by

\begin{align}
\chi_{\rm hex}=
\begin{pmatrix}
	\chi_{xx} & 0 & 0\\
	0 & \chi_{xx}& 0\\
	0 & 0 & \chi_{zz}\\
\end{pmatrix}.
\end{align}\\
Therefore, as long as the $C_6$ symmetry is preserved, the leading term of the torque is exactly zero. At high temperatures, CsV$_3$Sb$_5$ has a hexagonal crystal structure, so we expect $\tau(\theta,\phi)=0 $ as long as ${\bm M}$ is proportional to ${\bm H}$. 

If the in-plane rotational symmetry is broken and the crystal becomes orthorhombic, $\chi_{yy}$ is no longer equal to $\chi_{xx}$, and the susceptibility tensor is given by Eq.\,(\ref{ort}). Note that the $x$-axis defined along the nematic direction. In this case, $\Delta \chi = \chi_{xx}-\chi_{yy}$ is finite, and the sinusoidal response appears in the angular dependence of $\tau(\theta,\phi)$. Such a signal from the nematic state has also been observed in cuprates, irridates, iron-based superconductors and URu$_2$Si$_2$~\cite{murayama2019diagonal,murayama2021bond,kasahara2012electronic,okazaki2011rotational}.\\

\noindent
{\bf Time-reversal-symmetry breaking}
\noindent

The above discussion is based on the assumption that the sample is paramagnetic, i.e., ${\bm M}$ is linearly coupled to the magnetic field: ${\bm M} = \chi{\bm H}$. Provided this condition, the angular dependence of the magnetic torque response $\tau(\theta,\phi)$ is continuous and sinusoidal. However, $\tau(\theta,\phi)$ can show cusps, discontinuous jumps, and hysteresis loops if the system is either ferromagnetic, antiferromagnetic, or superconducting. In particular, jumps are often observed when the magnetic field passes across the easy axis and the spin texture of the sample is flipped. For example, jumps and cusps are observed in ferromagnetic VI$_3$~\cite{koriki2021magnetic}, antiferromagnetic Sr$_2$IrO$_4$~\cite{fruchter2016magnetic} and superconducting CaFe$_{0.88}$Co$_{0.12}$AsF~\cite{xiao2016angular}, but $\tau(\theta,\phi)$ becomes sinusoidal curve as these systems approach paramagnetic states. Therefore, our observation of distinct jumps and hysteresis loop strongly resembles the magnetic state, suggesting the broken time-reversal symmetry.\\\\

\begin{figure}[h]
	\label{fig:TRSB2}
	\includegraphics[clip,width=0.5\linewidth]{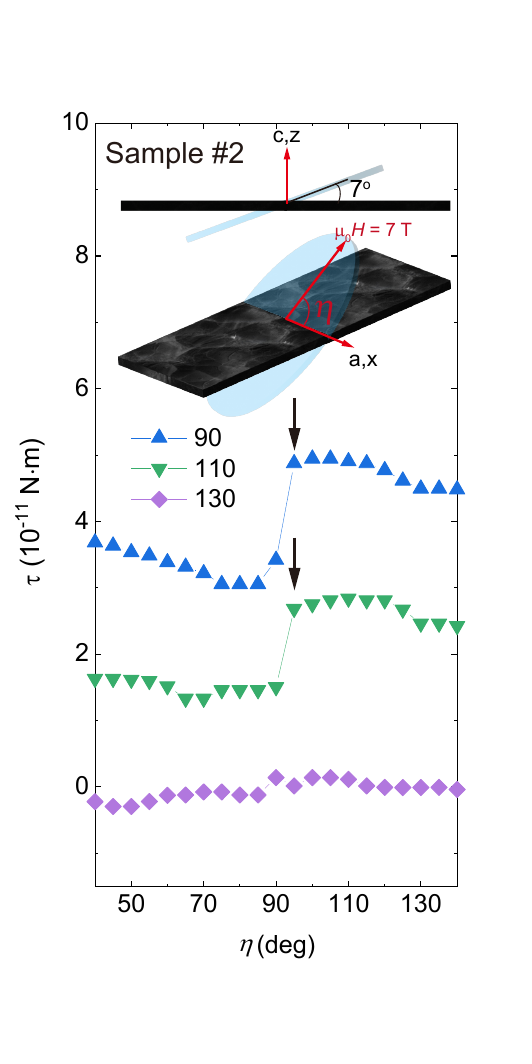}
	\caption{\textbf{Broken time-reversal symmetry from sample \#2.} The magnetic torque $\tau$($\eta$) of sample \#2 in the rotated magnetic field of $|\mu_0H|=$7\,T. The magnetic field is rotated at a tilting angle of 7$^\circ$ with respect to $ab$ plane, as illustrated in the inset.  We define the rotation angle $\eta$ from $a$-axis. Arrows indicate the first order transition. 
	}
\end{figure}

\noindent
{\bf First-order phase transition induced by $c$-axis field in sample \#2}

We confirmed the first-order phase transition induced by the $c$ axis component of ${\bm H}$ in sample \#2, as shown in Fig.\,S3.  As sample \#2 is much smaller than \#1, we rotated the magnetic field of 7\,T. The magnetic field is rotated at a tilting angle of 7$^\circ$ with respect to $ab$ plane, as illustrated in the inset. As indicated by arrows, we observe the first order transitions  below $T^*$, which are also observed in Fig\,4\FigCap{a} and \FigCap{b}.  The first order transition is induced at $\eta\approx 90^{\circ}$, where $c$ axis component of $H$ is maximum.
